\documentclass[floats,floatfix,amssymb,prd,twocolumn,superscriptaddress,nofootinbib]{revtex4-2}

\usepackage{amssymb,amsmath,verbatim,mathtools,needspace,enumitem,etoolbox,graphicx,physics,microtype,afterpage,bm}
\usepackage[dvipsnames, usenames]{xcolor}
\definecolor{linkcolor}{rgb}{0.0,0.3,0.5}
\usepackage[unicode, colorlinks=true, linkcolor=linkcolor, citecolor=linkcolor, filecolor=linkcolor,urlcolor=linkcolor, pdfusetitle]{hyperref}
\usepackage[all]{hypcap}
\usepackage[T1]{fontenc}
\usepackage[utf8]{inputenc}
\usepackage{tabularx}
\usepackage{float}
\usepackage{lmodern}

\allowdisplaybreaks

\usepackage{tikz}
\usepackage{color}
\usepackage{framed}
\usepackage{hyperref}
\hypersetup{colorlinks, citecolor=bluscuro, linkcolor=black, urlcolor=bluscuro}

\usepackage{bm}
\usepackage{latexsym}
\usepackage{amsmath}
\usepackage{amssymb}
\usepackage{amsthm}
\usepackage{amscd}
\usepackage[mathscr]{eucal}
\usepackage{graphicx}
\usepackage{subfigure}
\usepackage{psfrag}
\usepackage{rotating}

\usepackage{color}

\definecolor{ggreen}{cmyk}{0.7,     0,      0.9,      0}
\definecolor{viol}{cmyk}{0.3,1,0,0}
\definecolor{myred}{cmyk}{0.1, 1, 0.5, 0}
\definecolor{bblue}{rgb}{0.2, 0.29996, 0.8 }

\definecolor{rossos}{cmyk}{0,1,1,0.55}
\definecolor{bluscuro}{rgb}{0.15, 0.2, .85}
\definecolor{bluchiaro}{cmyk}{1,.3,0.,0.1}

\definecolor{ForestGreen}{rgb}{0.13, 0.55, 0.13}

\theoremstyle{plain}

\begin{document}


\title{Compactness bounds in General Relativity}

\author{Artur Alho}
\affiliation{Center for Mathematical Analysis, Geometry and Dynamical Systems, Instituto Superior T\'ecnico, Universidade de Lisboa, Av. Rovisco Pais, 1049-001 Lisboa, Portugal}

\author{Jos\'e Nat\'ario}
\affiliation{Center for Mathematical Analysis, Geometry and Dynamical Systems, Instituto Superior T\'ecnico, Universidade de Lisboa, Av. Rovisco Pais, 1049-001 Lisboa, Portugal}

\author{Paolo Pani}
\affiliation{Dipartimento di Fisica, Sapienza Universit\`a di Roma \& INFN Roma1, Piazzale Aldo Moro 5, 00185, Roma, Italy}

\author{Guilherme Raposo}
\affiliation{CENTRA, Instituto Superior T\'ecnico, Universidade de Lisboa, Av. Rovisco Pais, 1049-001 Lisboa, Portugal}
\affiliation{Centre for Research and Development in Mathematics and Applications (CIDMA), Campus de Santiago, 3810-183 Aveiro, Portugal}


\begin{abstract}

A foundational theorem due to Buchdahl states that, within General Relativity~(GR), the maximum compactness $\mathcal{C}\equiv GM/(Rc^2)$ of a static, spherically symmetric, perfect fluid object of mass $M$ and radius $R$ is $\mathcal{C}=4/9$. As a corollary, there exists a compactness gap between perfect fluid stars and black holes (where $\mathcal{C}=1/2$).
Here we generalize Buchdahl's result by introducing the most general equation of state for elastic matter with constant longitudinal wave speeds and apply it to compute the maximum compactness of regular, self-gravitating objects in GR.
We show that:~(i) the maximum compactness grows monotonically with the longitudinal wave speed; 
(ii) elastic matter can exceed Buchdahl's bound and reach the black hole compactness $\mathcal{C}=1/2$ continuously; (iii) however, imposing subluminal wave propagation lowers the maximum compactness bound to $\mathcal{C}\approx0.462$, which we conjecture to be the maximum compactness of \emph{any} static elastic object satisfying causality; (iv) imposing also radial stability further decreases the maximum compactness to $\mathcal{C}\approx 0.389$. 
Therefore, although anisotropies are often invoked as a mechanism for supporting horizonless ultracompact objects, we argue that the black hole compactness cannot be reached with physically reasonable matter within GR and that true black hole mimickers require either exotic matter or beyond-GR effects.   
\end{abstract}

\maketitle


{\it What is the maximum compactness that a self-gravitating material object can support within Einstein's theory of General Relativity~(GR)?}
This fundamental question was addressed by H.\ A.\ Buchdahl in 1959 in the context of perfect fluid models~\cite{Buchdahl:1959zz}. He showed that self-gravitating,
isotropic (or mildly anisotropic, as was later proved~\cite{Andreasson:2007ck,Karageorgis:2007cy,Urbano:2018nrs}), spherically symmetric, perfect fluid GR solutions satisfy the following bound on the compactness: $\mathcal{C}\equiv GM/(Rc^2)\leq4/9\approx 0.444$, where $M$ and $R$ are the object's mass and radius.
On the other hand, the horizon radius of a Schwarzschild black hole~(BH) is $R=2GM/c^2$, i.e. $\mathcal{C}=1/2$. Thus, as an important consequence, Buchdahl's theorem forbids the existence within GR of fluid objects whose compactness is arbitrarily close to the BH limit, providing an important cornerstone for tests of the nature of compact objects~\cite{Cardoso:2019rvt}.

In this letter we extend Buchdahl's foundational result to the case of elastic matter, unveiling several interesting features of extreme compact objects in GR.
Henceforth we set the gravitational constant $G$ and the speed of light $c$ to unity.

%

We focus on spherical symmetry and study static, self-gravitating configurations of elastic matter following the novel formalism developed in Refs.~\cite{Alho:2021sli,inprep}. 
The matter sector is described by the stress-energy tensor ${T^{\nu}_{\mu}={\rm diag}(\rho, p_\mathrm{rad}, p_\mathrm{tan},p_\mathrm{tan})}$, in terms of the density $\rho(r)$ and radial and tangential pressures $p_\mathrm{rad}(r)$ and $p_\mathrm{tan}(r)$, respectively. The ansatz for the geometry's line element reads ${ds^2=-e^{2\alpha(r)}dt^2+dr^2/(1-2m(r)/r)+r^2 d\Omega^2}$, where ${d\Omega^2}$ is the metric of the unit $2$-sphere.
The metric functions satisfy the usual relations $dm/dr=4\pi r^2 \rho$ and $ d\alpha/dr= 2\left({m}+{4\pi  r^3}p_\mathrm{rad}\right)/(r(r-2m))$, whereas the radial pressure satisfies a modified Tolman-Oppenheimer-Volkoff equation,
\begin{equation}
    \frac{dp_\mathrm{rad}}{dr} =\frac{2}{r}(p_\mathrm{tan}-p_\mathrm{rad})-(p_\mathrm{rad}+\rho)\frac{d\alpha}{dr}\,.\label{TOVeq} \\
\end{equation}
The system of field equations is closed by specifying an equation of state~(EoS) that relates the density and the pressures.
%
The latter can be imposed by introducing a stored energy function $w(r)$ that fully describes the properties of the elastic matter~\cite{Alho:2021sli,inprep}. The stored energy function can be conveniently written as a functional 	$\widehat{w}(\delta,\eta)$ that depends on two positive radial functions,
\begin{equation}
    \delta(r)=\frac{n(r)}{n_0}\,,\quad \eta(r)=\frac{3}{r^3}\int^{r}_{0}\frac{\delta(u)u^2du}{\left(1-{2m(u)}/{u}\right)^{1/2}}\,,
\end{equation}
related to the volumetric change and the stretches along the tangential direction, respectively. In the above equation, $n(r)$ is the baryon density and $n_0$ the baryon density in the reference (i.e., unstretched) state. Density, radial pressure, and tangential pressure can be obtained from
\begin{subequations}
\begin{align}
    \widehat{\rho}(\delta,\eta)&= \delta (\rho_0+ \widehat{w}(\delta,\eta))\,, \\
    \widehat{p}_\mathrm{rad}(\delta,\eta)&=\delta^2 \partial_\delta \widehat{w}(\delta,\eta) \,, \\
    \widehat{p}_\mathrm{tan}(\delta,\eta) &= \widehat{p}_\mathrm{rad}(\delta,\eta) + \frac32 \delta \eta \partial_\eta \widehat{w}(\delta,\eta)\,,
\end{align}\label{eq:EoS}
\end{subequations}
\noindent
which provide the desired EoS in parametric form (here $\rho_0>0$ is the density in the reference state).
The TOV equations for the stellar structure are solved by imposing regularity of the matter and metric functions at the center of symmetry, which implies {${(\delta(0),\eta(0))=(\delta_c,\delta_c)}$}. The radius $R$ of the star is defined by $p_\mathrm{rad}(R)=0$ and has a geometric meaning as the proper circumferential radius.

At variance with the fluid case, elastic materials are generically anisotropic and, in addition to the usual longitudinal matter perturbations, there exist also transverse perturbations.
For a spherical symmetric configuration, waves propagating along the radial direction can be longitudinal or transverse (we denote their speed by $c_\mathrm{L}(\delta,\eta)$ and $c_\mathrm{T}(\delta,\eta)$, respectively), whereas waves propagating along the tangential direction can be longitudinal (with speed $\tilde{c}_\mathrm{L}(\delta,\eta)$) or transverse along two orthogonal directions (with speeds $\tilde{c}_\mathrm{T}(\delta,\eta)$ and $\tilde{c}_\mathrm{TT}(\delta,\eta)$, respectively)~\cite{inprep}.

Within this general class of materials (that includes perfect fluids in the isotropic limit, $p_\mathrm{rad}=p_\mathrm{tan}\equiv p_\mathrm{iso}$, $c_\mathrm{L}=\tilde{c}_\mathrm{L}\equiv c_\mathrm{s}$, and $c_\mathrm{T}=\tilde{c}_\mathrm{T}=\tilde{c}_\mathrm{TT}=0$), our goal is to determine those which allow for the most compact self-gravitating solutions to Einstein's equations, while respecting causality.
In the fluid case, this can be achieved by requiring that the (adiabatic) sound speed $c_\mathrm{s}=\sqrt{{dp_\mathrm{iso}}/{d \rho}}$ be equal to the speed of light everywhere within the object, which singles out the EoS $p_\mathrm{iso}=\rho+{\rm constant}$, describing Christodoulou's hard phase fluid~\cite{Christodoulou:1995}.
In the general elastic case, we can consider the materials with longitudinal speeds of sound equal to the speed of light given in~\cite{Karlovini:2004gq,Natario:2019nrf}. In fact, the most general class of elastic materials with constant longitudinal sound speeds $c_\mathrm{L}=\tilde c_\mathrm{L}=\sqrt{\gamma-1}$ are given by the stored energy function~\cite{Karlovini:2004gq,inprep}
\begin{align}
\frac{\widehat{w}(\delta,\eta)}{\rho_0}&=  \left(\frac{1}{\gamma }-(2 \theta +\epsilon )\right)\delta ^{\gamma -1}+\delta^{-1}\left(\frac{\gamma-1}{\gamma}-(\theta +2 \epsilon )\right)\nonumber\\
&+\epsilon \frac{ \eta ^{\frac{\gamma}3}}{\delta } \left(2+\left(\frac{\delta}{\eta}\right)^{\gamma }\right)+\theta  \frac{\eta ^{\frac{2 \gamma }{3}}}{\delta} \left(1+2\left(\frac{\delta}{\eta}\right)^{\gamma}\right)-1\,,
\end{align}
depending on two dimensionless parameters $(\epsilon,\theta)$ (and also on the adiabatic index $\gamma>1$, with $\gamma\leq2$ imposed by subluminality).
The quantities $\rho_0$, $\epsilon$, and $\theta$ can be related to the Lam\'e parameters $\lambda$, $\mu\geq0$, and the Poisson ratio $\nu\in(-1,\frac{1}{2}]$ by 
\begin{equation}
 \rho_0=\frac{\lambda+2\mu}{\gamma-1}, \quad \epsilon+\theta=\frac{\gamma-1}{\gamma^2}\frac{2\mu}{\lambda+2\mu}=\frac{(\gamma-1)}{\gamma^2}\frac{(1-2 \nu)}{(1-\nu)}.  
\end{equation}
The isotropic fluid limit $(\epsilon,\theta)\to(0,0)$ ($\nu\to1/2$) of this family corresponds to a linear EoS, $\widehat{p}_\mathrm{rad}(\delta)=\widehat{p}_\mathrm{tan}(\delta)=\widehat{p}_\mathrm{iso}(\delta)=c_\mathrm{s}^2 (\widehat{\rho}(\delta)-\rho_0)$, where $c_\mathrm{s}=c_{\mathrm{L}}=\tilde{c}_{\mathrm{L}}$ is the (adiabatic) sound speed of the fluid, and $c_\mathrm{T}=\tilde{c}_\mathrm{T}=\tilde{c}_\mathrm{TT}=0$ as expected. The incompressible fluid limit which saturates the Buchdahl bound~\cite{Buchdahl:1959zz,Schwarzschild:1916} corresponds to $(\epsilon,\theta)\to(0,0)$ and $c_\mathrm{s}\to\infty$. The SUREOS material studied in~\cite{Karlovini:2004gq} corresponds to $\gamma=2$ (that is, $c_{\mathrm{L}}=\tilde{c}_{\mathrm{L}}=1$) and $(\epsilon,\theta)=(0,\frac14)$; for later comparison, we note that the maximum compactness found in~\cite{Karlovini:2004gq} for SUREOS stars (with a suitable material metric) was $\mathcal{C}\simeq 0.401$, decreasing to $\mathcal{C}\simeq 0.341$ when considering only radially stable stars.


\begin{figure}[t]
	\centering
	\label{fig:Ceps_examples}
	\includegraphics[width=0.48\textwidth]{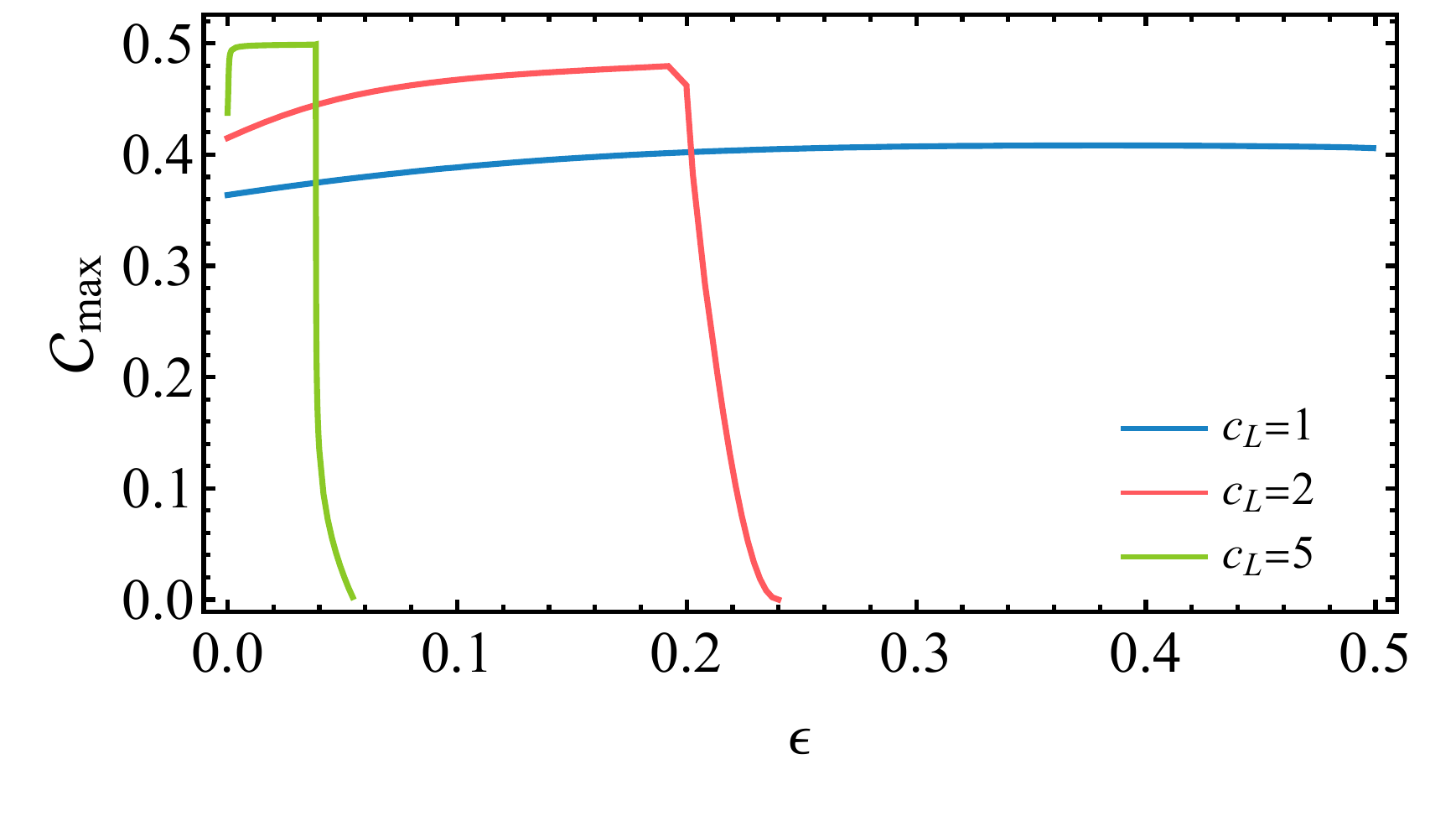} 
	\caption{Maximum compactness of static, self-gravitating elastic materials as a function of the parameter $\epsilon$ for $\theta=0$ and different values of longitudinal wave speeds $c_\mathrm{L}=\tilde c_\mathrm{L}$. The fluid limit corresponds to $\epsilon=0$.
	}
\end{figure}

For a given choice of the parameters $(\rho_0,\epsilon,\theta,\gamma)$, we obtain a one-parameter family of solutions in terms of ${\delta(0)=\delta_c}$, or, equivalently, in terms of the central density $\rho_c=\widehat{\rho}(\delta_c,\delta_c)$. In fact, $\rho_0$ is simply a scale factor which does not affect dimensionless quantities like the compactness.
For each of these families, as we vary the central density we find that the compactness is bounded.
Since the parameter space is multidimensional, for simplicity in Fig.~\ref{fig:Ceps_examples} we show the maximum compactness for $\theta=0$ families of solutions as a function of $\epsilon$  and some representative values of $c_\mathrm{L}$.
%
%
As evident from Fig.~\ref{fig:Ceps_examples}, the maximum compactness grows with the longitudinal sound speed. In the $(\epsilon,\theta)\to(0,0)$ limit we recover Buchdahl's bound, $\mathcal{C}_{\rm max}=4/9$, which is obtained for infinite sound speed. Interestingly, even for small deformations away from the fluid limit, Buchdahl's bound abruptly tends to the BH compactness, i.e. 
\begin{equation}
 \mathcal{C}_{\rm max}\to 1/2 \quad {\rm as} \quad c_\mathrm{L}\gg1 
\end{equation}
with $(\epsilon,\theta)\neq(0,0)$.
As an aside, the sharp decrease in the maximum compactness for configurations with ${c_L>1}$ (red and green lines of  Fig.~\ref{fig:Ceps_examples}) is due to the central pressure becoming a decreasing function of $\delta_c$ for sufficiently large $\epsilon$; the maximum compactness becomes zero when there is no positive central pressure for any $\delta_c>1$.


It is worth noting that the BH limit, $\mathcal{C}\to1/2$, is obtained in the same unphysical configuration as the Buchdahl limit for fluids, namely when the wave propagation speed diverges. A more relevant question concerns the maximum compactness of \emph{physically admissible}~(PA) solutions. The latter require real subluminal wave propagation speeds,
\begin{equation}\label{eq:causal}
    c_\mathrm{L}, \tilde{c}_\mathrm{L}, c_\mathrm{T}, \tilde{c}_\mathrm{T}, \tilde{c}_\mathrm{TT}\leq 1\,,
\end{equation}
and that all energy conditions hold~\cite{HawkingEllis}. We have addressed this question through a detailed numerical investigation of the entire parameter space.  
In particular, for each point of the $(\epsilon,\theta)$ plane, we have computed a one-parameter family of solutions, checking subluminality of all sound speeds and the energy conditions within the object.
This is particularly challenging in this model, because in some regions of the parameter space subluminality and energy conditions might be violated for some density ranges but not in general, as discussed below.
We found that the maximum compactness of PA elastic solutions occurs for $\epsilon + \theta<0$ and $\theta>0$, and that it saturates to the bound 
\begin{equation}
    \mathcal{C}_{\rm PA}\lesssim 0.462\,, \label{eq:PA}
\end{equation}
as the elastic parameters become sufficiently large in absolute value. Interestingly, this bound is sufficiently large as to allow for stars with a photon sphere ($\mathcal{C}>1/3$), and it is even larger than the (unphysical) Buchdahl limit in the fluid case. It is worth noting that  by imposing causality in the fluid limit we recover the so-called \emph{causal} Buchdahl bound~\cite{Urbano:2018nrs,Boskovic:2021nfs}, $\mathcal{C}_{\rm PA}^{\rm fluid}\lesssim 0.364$, which is much smaller than the bound in Eq.~\eqref{eq:PA}.

\begin{figure}[t]
	\centering
	\label{fig:MR}
	\includegraphics[width=0.48\textwidth]{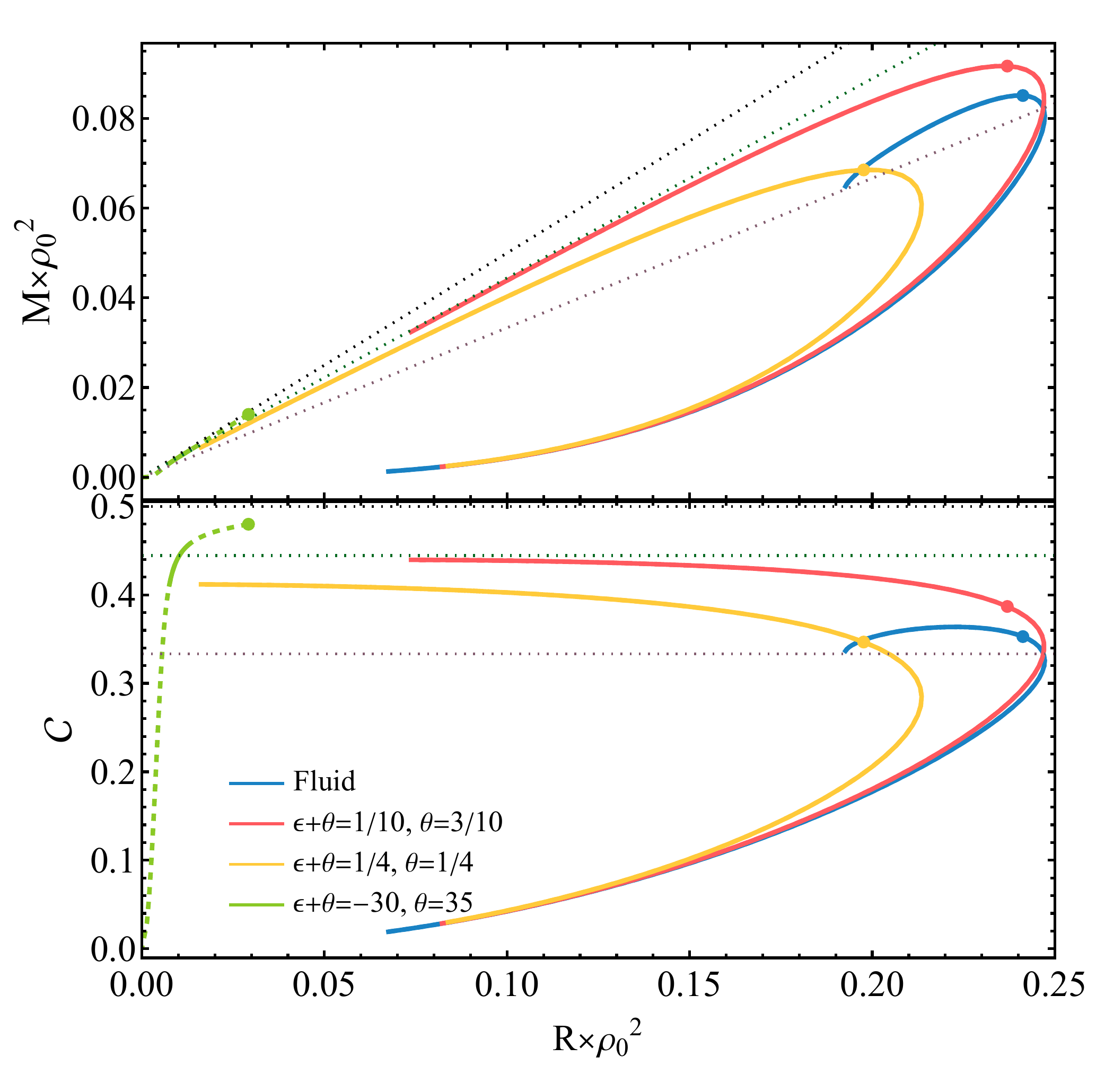} \\

	\caption{Mass-radius (top panel) and compactness-radius (bottom panel) diagrams for some elastic solutions saturating the causality condition, $c_\mathrm{L}=\tilde{c}_\mathrm{L}=1$. Continuous (dashed) curves correspond to the physical (unphysical) branch of the solutions. The markers on each curve represent the maximum mass configuration. The configurations on the right of the marker belong to the radial stable branch, whereas configurations on the left are radially unstable. For the green curve all configurations are radially unstable. 	}
\end{figure}

In Fig.~\ref{fig:MR} we show the mass-radius and compactness-radius diagrams for some families of solutions with $c_\mathrm{L}=\tilde{c}_\mathrm{L}=1$: a perfect-fluid solution (blue curve), a solution which yields the most compact and stable configuration (red curve, see below for the stability analysis), the SUREOS material~\cite{Karlovini:2004gq} (yellow), and a solution that saturates the bound~\eqref{eq:PA} (green). While the first two elastic solutions (red and yellow) are qualitatively similar to the fluid case and contain only PA configurations for all densities that we probed, the green curve is qualitatively very different. This difference is associated to the fact that $\epsilon+\theta<0$ for the green curve, which brings the parameters of linear elasticity outside their usual physical range. 
However, since the reference state of astrophysical compact objects is not observable, as they only exist in their deformed state, we are not constrained by this requirement. In fact, the green curve exhibits two independent branches which are unphysical for most values of the central density. The first unphysical branch (not shown in Fig.~\ref{fig:MR}) corresponds to lower central densities, and as the central density increases the energy density of the star can become negative (thus violating the weak energy condition), eventually leading to negative stellar masses. For high enough values of the central density, the solutions enter a new branch (the one shown in Fig.~\ref{fig:MR}), which, although composed mostly of unphysical configurations (dashed curve), also contains a small region where all the energy conditions are satisfied and all velocities are subluminal (solid curve).

\begin{figure}[t]
	\centering
	\label{fig:Ceps}
	\includegraphics[width=0.48\textwidth]{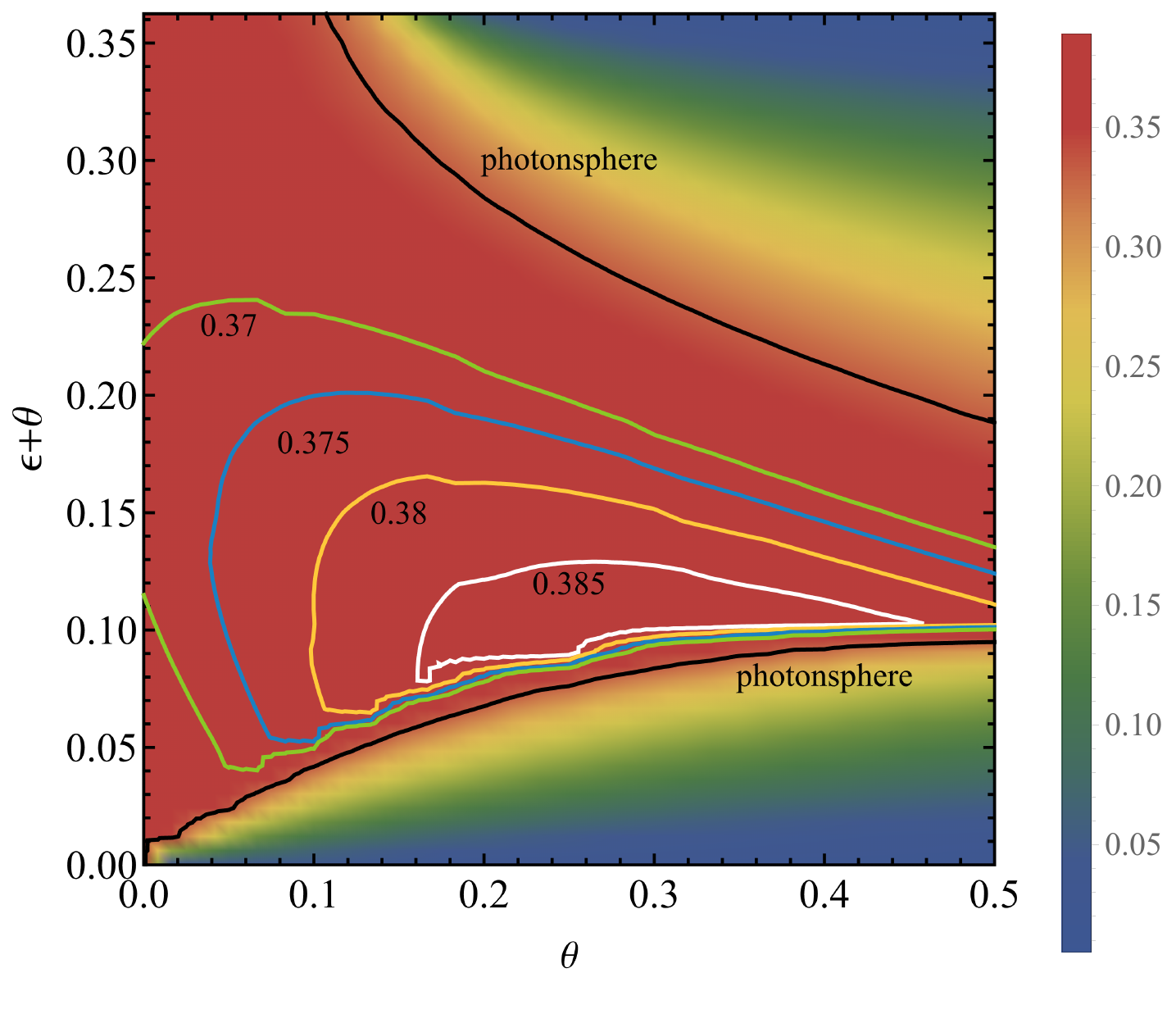} 
	\caption{Density plot for the maximum compactness of physically admissible (all wave speeds are causal and energy conditions are satisfied) and radially stable solutions in the two-dimensional parameter space ($\theta,\epsilon+\theta$).
	}
\end{figure}

A further condition for the physical admissibility of the solutions is to require stability of the equilibrium configurations. In the fluid case this constraint is more stringent than causality and imposes $\mathcal{C}_{\rm PAS}^{\rm fluid}\lesssim 0.354$ on physically admissible, radially stable~(PAS) configurations~\cite{Lindblom:1984,Lattimer:2006xb,Urbano:2018nrs,Boskovic:2021nfs}.
Similarly to perfect fluids, elastic stars are radially stable (unstable) for densities below (above) that of corresponding maximum mass~\cite{Alho:2021sli,inprep}. In the mass-radius and compactness radius diagrams in Fig.~\ref{fig:MR}, this property translates into two branches, one stable and one unstable, respectively on the right and left of the maximum mass configuration (circular marker along each curve).  

In Fig.~\ref{fig:Ceps} we show the density plot for the maximum compactness of PAS configurations.
Overall, the maximum compactness obtained by requiring radial stability is always more stringent than that obtained from causality alone for each point of the parameter space.
The maximum compactness configuration that satisfies the physical admissibility conditions and at the same time is radially stable corresponds to $(\epsilon,\theta)\sim (-0.2,0.3)$. This provides the bound
\begin{equation}
    \mathcal{C}_{\rm PAS}\lesssim 0.389\, , \label{boundPAS}
\end{equation}
which is significantly larger than in the fluid limit (even when relaxing the stability requirement for fluids).

\begin{figure}[t]
	\centering
	\label{fig:DEC}
	\includegraphics[width=0.48\textwidth]{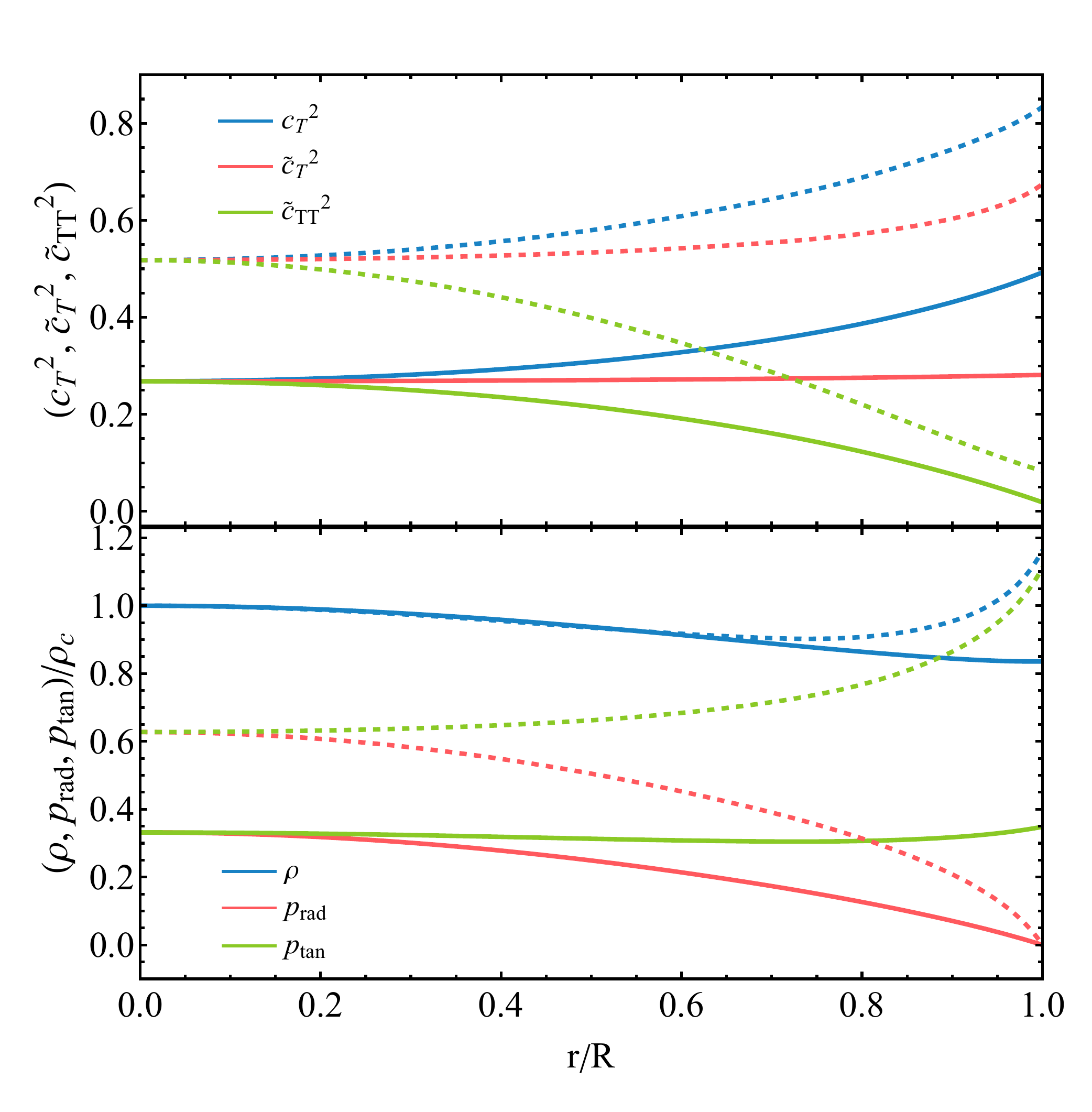} 
	\caption{Wave speed (upper panel) and matter density and pressure profiles (bottom panel) for solutions saturating the ${\cal C}_{\rm PAS}$ bound (solid curves) and  the ${\cal C}_{\rm PA}$ bound (dashed curves). 
	All non-trivial wave speeds are well-defined and subluminal, and the energy conditions are satisfied. 	}
\end{figure}

Finally, to illustrate the physically admissible conditions of the solutions, in Fig.~\ref{fig:DEC} we show the non-trivial sound speed profiles and the density, radial and tangential pressures for two configurations with $c_\mathrm{L}=\tilde{c}_\mathrm{L}=1$ which approach either the~PAS bound (solid lines) or the less stringent PA bound (dashed lines).
Furthermore, as shown in the example of Fig.~\ref{fig:DEC}, physically admissible solutions are also regular and smooth in the stellar interior.

%

%

To summarize, we have shown that Buchdahl's bound for perfect fluids in GR, $\mathcal{C}\leq 4/9$, is generically exceeded in the presence of elasticity. At variance with the fluid case, the maximum compactness can be arbitrarily close to the BH value, $\mathcal{C}=1/2$. However, this occurs only for superluminal wave propagation in the material. Physically admissible solutions have $\mathcal{C}\lesssim 0.462$, which is further lowered to $\mathcal{C}\lesssim 0.389$ by additionally requiring radial stability.
This also generalizes our previous findings~\cite{Alho:2021sli}, confirming that physically admissible, radially stable elastic stars can be surrounded by a photon sphere at $r=3M$, and might therefore feature gravitational-wave echoes~\cite{Cardoso:2016rao,Cardoso:2017cqb,Pani:2018flj}. 
However, we have studied the dynamics of linear perturbations in a solution that almost saturates the bound~\eqref{boundPAS}, finding that even in this case the compactness is not sufficient to effectively confine quasibound states in the stellar interior~\cite{Cardoso:2019rvt}, so echoes are not efficiently produced. On the other hand, more compact PA solutions, which violate Eq.~\eqref{boundPAS} and are radially unstable, do feature echoes.
An interesting open question is whether PAS solutions are prone to nonlinear instabilities possibly associated with quasi-bound, long-lived modes trapped between the photon sphere and the stellar interior~\cite{Cardoso:2014sna,Keir:2014oka,Cunha:2017qtt,Ghosh:2021txu}.

While it is reasonable to expect that the most compact configurations are achieved when $c_\mathrm{L}=\tilde{c}_\mathrm{L}=1$ throughout the star, a natural question is what happens for variable wave propagation speeds within the object. We address this point in a forthcoming companion paper~\cite{inprep}, confirming the expectation that the $c_\mathrm{L}=\tilde{c}_\mathrm{L}=1$ model provides the largest compactness.
Based on our results, we conjecture that Eq.~\eqref{eq:PA} sets the maximum compactness for \emph{any} physically admissible, static, elastic object in GR. Future work could (dis)prove this conjecture. 

The existence of a maximum compactness smaller than $\mathcal{C}=1/2$ for physically admissible self-gravitating objects in GR implies that BHs form a discontinuous family of solutions. 
Reversing the argument, an independent measurement of the compactness exceeding the upper bound derived above --~as possibly achievable with gravitational wave observations~\cite{Cardoso:2019rvt,Maggio:2020jml,LIGOScientific:2021sio}, with the Event Horizon Telescope~\cite{EventHorizonTelescope:2019ggy,EventHorizonTelescope:2019pgp}, or other electromagnetic probes~\cite{Bambi:2015kza}~-- would either give further confirmation that the object is a BH or imply a violation of GR. 
Indeed, our analysis suggests that ultracompact objects that have been advocated as BH mimickers~\cite{Cardoso:2019rvt,Carballo-Rubio:2018jzw} might generically suffer from violations of the energy conditions or superluminal wave propagation when $\mathcal{C}\approx 1/2$, at least in the context of GR (explicit examples are gravastars~\cite{Mazur:2004fk,Pani:2015tga}, wormholes~\cite{Damour:2007ap}, anisotropic stars~\cite{Yagi:2015upa,Raposo:2018rjn,Biswas:2019gkw}, and others~\cite{Cardoso:2019rvt}). 

Our argument assumes staticity and spherical symmetry.
Like the fluid case~\cite{Stergioulas:2003yp}, we expect that including spin will increase the maximum compactness of the solutions by a small amount but will not change our main results qualitatively. The angular velocity of the object must be bounded by causality, so centrifugal effects are limited. In particular, we always expect a compactness gap between PAS stars and the corresponding BH family. Furthermore, in the spinning case the compactness of stable solutions should be further bounded by the ergoregion instability~\cite{1978CMaPh..63..243F,1978RSPSA.364..211C,Cardoso:2007az,Moschidis:2016zjy,Maggio:2017ivp,Maggio:2018ivz}, although the value of such an upper bound is probably model-dependent.

Finally, it is also interesting to note that compactness bounds may apply also to other kinds of matter fields. For example, minimally-coupled bosonic fields have an anisotropic stress-energy tensor and, in the case of strong self-interactions, can saturate the causal Buchdahl bound for fluids~\cite{Boskovic:2021nfs}. Whether the elastic Buchdahl bound discovered here can be exceeded by other forms of physically admissible matter is an intriguing open question.

\noindent{{\bf{\em Acknowledgments.}}}
%
We thank Alfredo Urbano for interesting discussion.
A.A.\ and J.N.\ were partially
supported by FCT/Portugal through CAMGSD, IST-ID, projects UIDB/04459/2020 and UIDP/04459/2020.
P.P.\ acknowledges financial support provided under the European Union's H2020 ERC, Starting 
Grant agreement no.~DarkGRA--757480, and under the MIUR PRIN and FARE programmes (GW-NEXT, CUP:~B84I20000100001), and support from the Amaldi Research Center funded by the MIUR program `Dipartimento di Eccellenza" (CUP:~B81I18001170001). G.R.\ was supported by the Center for Research and Development in Mathematics
and Applications (CIDMA) through the Portuguese Foundation for Science and Technology (FCT - Fundação para a Ciência e a Tecnologia), references UIDB/04106/2020 and UIDP/04106/2020.

\bibliographystyle{utphys}
\bibliography{biblio}

\end{document}